\documentclass[twocolumn]{aastex631}

\usepackage{graphicx}
\usepackage{float}
\usepackage{mathrsfs}	
\usepackage{amsmath}
\usepackage{color}
\usepackage{multirow}
\usepackage{subfigure}
\usepackage{longtable}
\usepackage{blindtext}
\usepackage{amssymb}
\def\gsim{~\rlap{$>$}{\lower 1.0ex\hbox{$\sim$}}}
\def\lsim{\mathrel{\rlap{\lower3.5pt\hbox{\hskip0.5pt$\sim$}}
    \raise0.5pt\hbox{$<$}}}

\shorttitle{AASTeX v6.3.1 Sample article}
\shortauthors{Li et al.}
\graphicspath{{./}{figures/}}

\begin{document}

\title{Optical+NIR analysis of a newly confirmed Einstein ring at z$\sim$1 from the Kilo-Degree Survey: \\Dark matter fraction, total and dark matter density slope and IMF

\footnote{Based on observations with OmegaCam@VST, VIRCAM@VISTA, HAWK-I and XSHOOTER@VLT (Prog. ID: 107.22S8).}}

\author[0000-0002-3490-4089]{Rui Li}
\affiliation{Institute for Astrophysics, School of Physics, Zhengzhou University, Zhengzhou, 450001, China.}

\author{Nicola R. Napolitano}
\affiliation{Department of Physics ``E. Pancini'', University Federico II, Via Cinthia 6, 80126-I, Naples, Italy}
\affiliation{School of Physics and Astronomy, Sun Yat-sen University, Zhuhai Campus, 2 Daxue Road, Xiangzhou District, Zhuhai, P. R. China.} 
\affiliation{CSST Science Center for Guangdong-Hong Kong-Macau Great Bay Area, Zhuhai, 519082, P. R. China}

\author[0000-0001-9697-7331]{Giuseppe D$'$Ago}
\affiliation{Institute of Astronomy, University of Cambridge, Madingley Road, Cambridge CB3 0HA, United Kingdom.}

\author{Vyacheslav N. Shalyapin}
\affiliation{Departamento de F\'\i sica Moderna, Universidad de Cantabria, Avda. de Los Castros s/n, E-39005 Santander, Spain}
\affiliation{O.Ya. Usikov Institute for Radiophysics and Electronics, National Academy of Sciences of Ukraine, 12 Acad. Proscury St., UA-61085 Kharkiv, Ukraine}

\author[0000-0002-2583-2669]{Kai Zhu}
\affiliation{Department of Astronomy, Tsinghua University, Beijing, Beijing 100084, China}

\author{Xiaotong Guo}
\affiliation{Institute of Astronomy and Astrophysics, Anqing Normal University, Anqing, Anhui 246133, China}

\author{Ran Li}
\affiliation{School of Physics and Astronomy, Beijing Normal University,  Beijing 100875, China.\\
School of Astronomy and Space Science, University of Chinese Academy of Science, Beijing 100049, China.}

\author{L\'{e}on V. E. Koopmans}
\affiliation{Kapteyn Astronomical Institute, University of Groningen, P.O.Box 800, 9700AV Groningen, the Netherlands}

\author{Chiara Spiniello}
\affiliation{Department of Physics, University of Oxford, Denys Wilkinson Building, Keble Road, Oxford OX1 3RH, UK}
\affiliation{INAF -- Osservatorio Astronomico di Capodimonte, Salita Moiariello 16, 80131 - Napoli, Italy}

\author{ Crescenzo Tortora}
\affiliation{INAF -- Osservatorio Astronomico di Capodimonte, Salita Moiariello 16, 80131 - Napoli, Italy}

\author{Francesco La Barbera}
\affiliation{INAF -- Osservatorio Astronomico di Capodimonte, Salita Moiariello 16, 80131 - Napoli, Italy}

\author{Haicheng Feng}
\affiliation{Yunnan Observatories, Chinese Academy of Sciences, Kunming, 650011, Yunnan, People's Republic of China}

\author{Liang Gao}
\affiliation{Institude for Astrophysics, School of Physics, Zhengzhou University, Zhengzhou, 450001, China}

\author{Zhiqi Huang}
\affiliation{School of Physics and Astronomy, Sun Yat-sen University, Zhuhai Campus, 2 Daxue Road, Xiangzhou District, Zhuhai, P. R. China}

\author{Koen Kuijken}
\affiliation{Leiden Observatory, Leiden University, P.O.Box 9513, 2300RA Leiden, The Netherlands}

\author{Hui Li}
\affiliation{Institude for Astrophysics, School of Physics, Zhengzhou University, Zhengzhou, 450001, China}

\author{ Linghua, Xie}
\affiliation{School of Physics and Astronomy, Sun Yat-sen University, Zhuhai Campus, 2 Daxue Road, Xiangzhou District, Zhuhai, P. R. China}

\author{Mario Radovich}
\affiliation{INAF - Osservatorio Astronomico di Padova, via dell'Osservatorio 5, 35122 Padova, Italy}

\author{Alexey Sergeyev}
\affiliation{Universit\'{e} C\^ote d’Azur, Observatoire de la C\^ote d’Azur, CNRS, Laboratoire Lagrange, France}
\affiliation{V. N. Karazin Kharkiv National University, Kharkiv, 61022, Ukraine}
\affiliation{Institute of Radio Astronomy of National Academy of Science of Ukraine, Mystetstv 4, Ukraine}



\begin{abstract}
We report the spectroscopic confirmation of a bright blue Einstein ring in the Kilo Degree Survey (KiDS) footprint: the Einstein ``blue eye''. Spectroscopic data from X-Shooter at the Very Large Telescope (VLT) show that the lens is a typical early-type galaxy (ETG) at $z_l=0.9906$, while the background source is a Ly$\alpha$ emitter at $z_s=2.823$. The reference lens modeling was performed on a high-resolution $Y-$band adaptive-optics image from HAWK-I at VLT. Assuming a singular isothermal ellipsoid (SIE) total mass density profile, we inferred an Einstein radius $R_{Ein}=10.47 \pm 0.06$ kpc. The average slope of the total mass density inside the Einstein radius, as determined by a joint analysis of lensing and isotropic Jeans equations is $\gamma_{tot}=2.14^{+0.06}_{-0.07}$, showing no systematic deviation from the slopes of lower redshift galaxies, This can be the evidence of ETGs developing through dry mergers plus moderate dissipationless accretion. Stellar population analysis with 8-band ($gri$ZYJHK$s$) photometries from KiDS and VIKING shows that the total stellar mass of the lens is $M*=(3.95\pm 0.35)\times 10^{11} M_\odot$ (Salpeter Initial Mass Function, IMF), implying a dark matter fraction inside the effective radius to be $f_{\rm DM}=0.307\pm 0.151$.
We finally explored the dark matter halo slope and found a strong degeneracy with the dynamic stellar mass. Dark matter adiabatic contraction is needed to explain the posterior distribution of the slope unless IMF heavier than Salpeter is assumed.

\end{abstract}

\keywords{Strong lensing --- galaxies --- Einstein ring}


\section{Introduction} \label{sec:intro}

Strong lensing is the effect of image distortion of faraway galaxies (sources) due to their light being bent by massive celestial bodies (lenses or deflectors), as predicted by general relativity. 
It is sensitive to the total mass, including dark matter (DM) and baryonic matter, and, as such, it is a crucial tool for studying the interplay between these two components
(\citealt{Koopmans2006ApJ...649..599K, Koopmans2009ApJ...703L..51K}; \citealt{Treu2004ApJ...611..739T, Treu2006ApJ...640..662T, Treu2011MNRAS.417.1601T}; \citealt{Auger2009ApJ...705.1099A}; \citealt{Bolton2012ApJ...757...82B}; \citealt{Sonnenfeld2013ApJ...777...98S, Sonnenfeld2014ApJ...786...89S}; 
\citealt{2020MNRAS.496.4717H};
\citealt{2023MNRAS.521.6005E};
\citealt{Li2018MNRAS.480..431L}). 
With strong lensing, we can measure mass with very high precision, typically with 5\% uncertainties (see a.g. \citealt{Bolton2008ApJ...682..964B}), whereas other methods, such as galaxy dynamics and X-rays, maybe face higher uncertainties (e.g., \citealt{Humphrey2006ApJ...639..136H, Napolitano2010MNRAS.405.2351N, Tortora2022A&A...657A..19T,Zhu2023MNRAS.522.6326Z}). More importantly, these latter techniques become less accurate as one goes to higher redshifts (see e.g. \citealt{Shetty2014ApJ...786L..10S}), where the strong lensing method is more efficient instead (see e.g. \citealt{Treu2004ApJ...611..739T}).

Strong lensing has been used as a probe of the $\Lambda$CDM predictions as it offers insights into the assembly history of galaxies and their relationship with their dark matter halo centers (e.g., \citealt{2001Ap&SS.276..851F, Dehne2005MNRAS.363.1057D, Kazantzidis2006ApJ...641..647K, Nipoti2009ApJ...703.1531N}), or to probe the slope of the total mass (dark plus stellar mass) density. Statistical analyses of strong lensing systems have consistently indicated that the total mass distribution of massive early-type galaxies (ETGs) follows a power-law profile, $\rho \propto r^{-\gamma}$, with $\gamma \approx 2$ and uncertainties below 10\% (\citealt{Koopmans2006ApJ...649..599K, Koopmans2009ApJ...703L..51K}, \citealt{Auger2009ApJ...705.1099A}, \citealt{Tortora2014MNRAS.445..115T}, \citealt{Sonnenfeld2013ApJ...777...98S}, \citealt{2023arXiv231109307T}), under the assumption of isotropic velocity dispersion.
This result was further confirmed by dynamics, showing an isothermal (e.g., \citealt{Tortora2016ASSP...42..215T}) or slightly steeper (e.g. $\gamma \sim 2.2$, \citealt{Bellstedt2018MNRAS.476.4543B, Zhu2023MNRAS.522.6326Z, Lishubo2023arXiv231013278L, Zhu2024MNRAS.527..706Z, Wang2024MNRAS.527.1580W}) total mass density slope.



Hydrodynamical simulations have shown that $\gamma\sim2$ is 
predicted at lower redshifts, where galaxy growth is driven by ``dry" mergers (e.g., \citealt{Nipoti2009ApJ...703.1531N}, \citealt{Remus2017MNRAS.464.3742R}). However, the 
redshift evolution of $\gamma$ 
is still debated, as simulations predict the slope to become steeper ($\gamma>2$) with increasing redshift (\citealt{Remus2017MNRAS.464.3742R}, \citealt{Wang2019MNRAS.490.5722W}), while some observations
indicate a rather constant (\citealt{Sonnenfeld2013ApJ...777...98S}), 
or decreasing $\gamma$ with redshift (\citealt{Bolton2012ApJ...757...82B, Li2018MNRAS.480..431L}). Unfortunately, most of these measurements are limited to $z \lsim 0.8$, with very few excursions to higher redshifts 
(MG 2016+112 at $z=1.004$, CFRS03.1077 at $z=0.938$; \citealt{Koopmans2002ApJ...568L...5K},
\citealt{Treu2002ApJ...575...87T, Treu2004ApJ...611..739T}).
Hence, this poor statistics of lenses at $z > 0.8$ is a serious handicap to resolving the controversy between observations and simulations.
\begin{figure*}
\centering
\includegraphics[width=\textwidth]{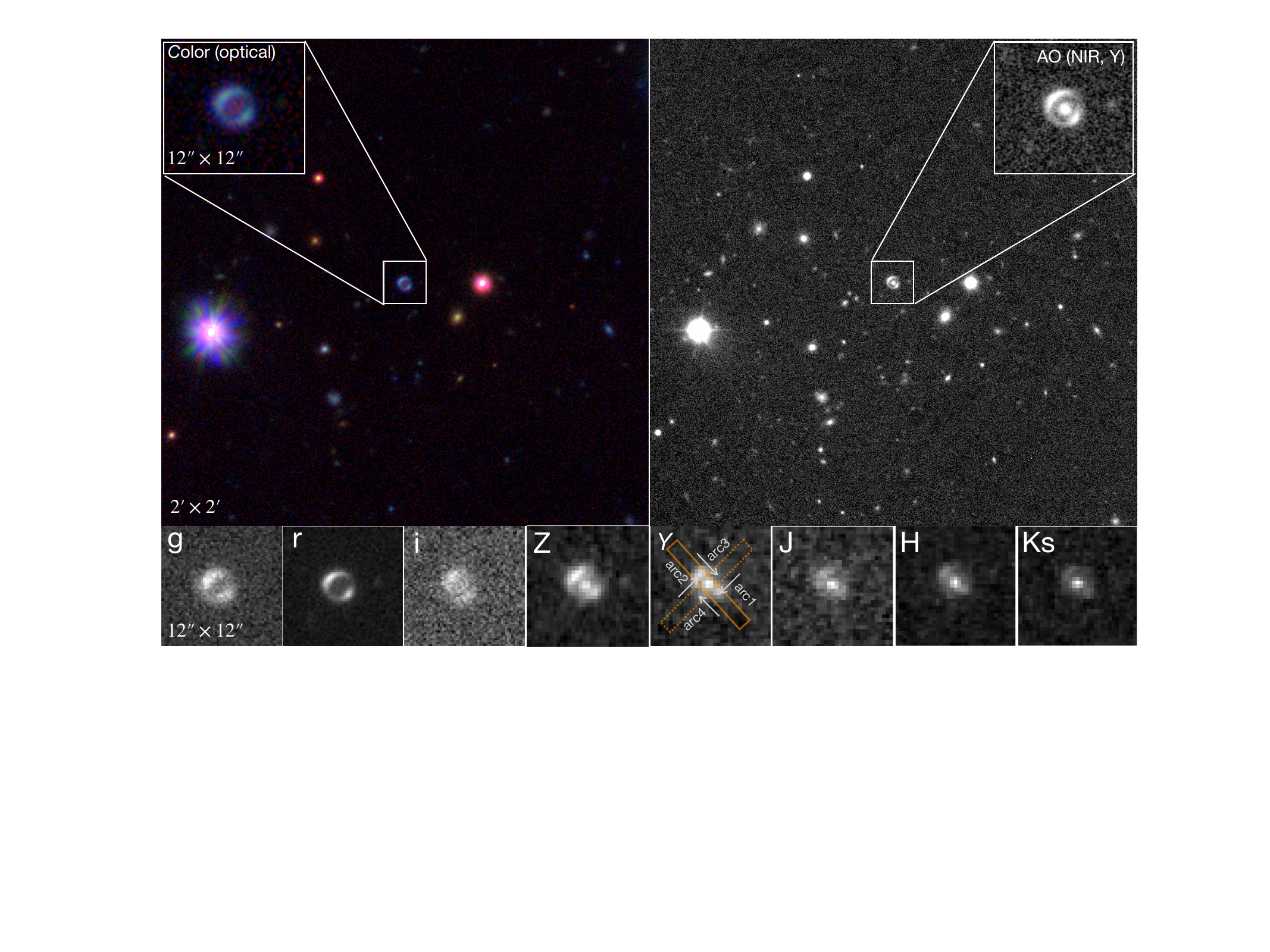}
\caption{Images of KIDSJ0233-3447. Top left: colour-composited images  ($2'\times2'$) from KiDS observation and the zoom-in of the lens  ($12''\times12''$). Top right: Y-band image from HAWKI@VLT ($2'\times2'$) and the zoom-in of the lens ($12''\times12''$). Bottom: 8-band images from KiDS and VIKING.
The two slits used for the spectroscopical observations are shown in orange dashed  (for lens) and solid (for arcs) rectangles on the Y-band image, respectively. White arrows labeled with arc 1 to arc 4 show the approximate location of the regions used to extract the spectra of the arcs (see Sec. \ref{sec:spectrum})}.
\label{fig:multi_band_images}
\end{figure*}

Another major source of systematics in the mass slope problem is the assumed stellar Initial Mass Function (IMF). This strongly impacts the stellar mass, $M_*$, and consequently the dark matter fraction $f_{\rm DM}$ of the foreground lens. Changes in the IMF normalization can increase the stellar mass-to-light ratio by more than a factor of 2, as evidenced by a large number of observations in massive elliptical galaxies
(\citealt{Conroy2012ApJ...760...71C}, \citealt{Cappellari2012Natur.484..485C},
\citealt{LaBarbera2013MNRAS.433.3017L},
\citealt{Tortora2013ApJ...765....8T},
\citealt{2014MNRAS.438.1483S}, \citealt{2017ApJ...838...77L}, \citealt{2023arXiv230912395L}).
The choice of the IMF significantly influences conclusions regarding the DM slope. Specifically, if the IMF leads to a higher estimate of stellar mass in the galaxy center, this implies that there is less DM present. Consequently, the DM density profile must be less steep to remain consistent with the observed stellar kinematics.
And vice versa for lighter IMF, which leads to lower stellar mass $M_*$, steeper slopes have to be assumed to replace the missing stellar masses. This can be reconciled with $\Lambda CDM$ only by invoking some form of adiabatic contraction (e.g., \citealt{Napolitano2010MNRAS.405.2351N}, \citealt{Tortora2010ApJ...721L...1T}, \citealt{Treu2010ApJ...709.1195T}). 
The degeneracy between IMF assumptions and DM slope estimates becomes particularly problematic when considering the growing evidence that the IMF may not be universal (\citealt{Cappellari2012Natur.484..485C}, \citealt{Tortora2013ApJ...765....8T}, \citealt{Spiniello2014MNRAS.438.1483S}, \citealt{2015ApJ...806L..31M},
\citealt{2021A&A...645L...1B}, 
\citealt{2023arXiv231213355M}, \citealt{2012Greggio_Renz_IMF_redshift}). Variations in IMF with galaxy mass, stellar population properties and redshift introduce additional systematic uncertainties, as different galaxies may require distinct IMF prescriptions.



In our previous work (\citealt{Li_DR5lens_2021}), we identified a blue Einstein ring candidate, KIDSJ0233-3447, which was also independently discovered by \citealt{Stein2022ApJ...932..107S}. 
Subsequent spectroscopic follow-up observations have confirmed this object to be a genuine strong lens with a foreground galaxy at redshift $z_l=0.9906$. For the peculiar appearance in the optical bands, we have dubbed it the Einstein ``blue eye'' (see Fig. \ref{fig:multi_band_images}).
KIDSJ0233-3447 is the second most distant lens for which such a detailed analysis (e.g., velocity dispersion, stellar mass, total mass density slope, dark matter fraction and its density slope) has been performed (MG 2016+112 mentioned above being the most distant one, \citealt{Koopmans2002ApJ...568L...5K}) and among only a limited number of confirmed lenses at $z_l>0.9$ (e.g.,
\citealt{Wong2014ApJ...789L..31W},
\citealt{Barone-Nugent2015MNRAS.453.3068B},
\citealt{Canameras2017A&A...600L...3C},
\citealt{Ciesla2020A&A...635A..27C}, \citealt{vanDokkum2024NatAs...8..119V}), hence providing us a unique opportunity to move a step toward the understanding of the mass properties of galaxies in their early evolution phases.

In this paper, we present a comprehensive analysis of the stellar and dark matter properties of KIDSJ0233-3447 and provide a detailed discussion of IMF and the profiles of its total and dark matter mass. The paper is organized as follows: in Section 2, we introduce the data analysis. In Section 3, we provide a discussion of the results on the total mass slope, dark matter properties, and IMF model. Finally, in Section 4, we summarize our findings and draw our conclusions.  For all calculations, we assume a $\rm \Lambda$CDM cosmology with the WMAP7 (\citealt{WMAP7_2011ApJS..192...18K}) results: ($\rm \Omega_\Lambda$, $\rm \Omega_M$, h)=(0.272, 0.728, 0.704).

\section{data analysis and key parameters}
\label{sec:data}
In order to perform a complete, combined and self-consistent lensing modeling, dynamics, and stellar population analysis on  KIDSJ0233-3447, we have collected seeing matched $gri$ZYJHK$_s$ photometry from the Kilo Degree Survey (KiDS, \citealt{deJong2013ExA....35...25D}, \citealt{2024A&A...686A.170W}) and VISTA Kilo-degree Infrared Galaxy (VIKING, \citealt{Edge2013Msngr_VIKING}) survey and a similarly wide wavelength baseline spectroscopy from X-Shooter, and the adaptive-optics (AO) NIR data from HAWK-I at VLT (see \S\ref{sec:data}). Moreover, these data also offer the chance to assess the importance of optical + NIR datasets in strong lensing analyses, for which the KiDS + VIKING dataset is the only available precursor of optical + NIR jointing analysis of Euclid (\citealt{Laureijs+11_Euclid}), 
Vera C. Rubin Observatory (\citealt{Izevic+19_LSST}), Roman (\citealt{2015arXiv150303757S_Roman}) and the China Space Station Telescope (\citealt{Zhan+18_csst}).

\subsection{Spectroscopic confirmation}
\label{sec:spectrum}

\begin{figure}
\hspace{-0.5cm}
\includegraphics[width=9.2cm]{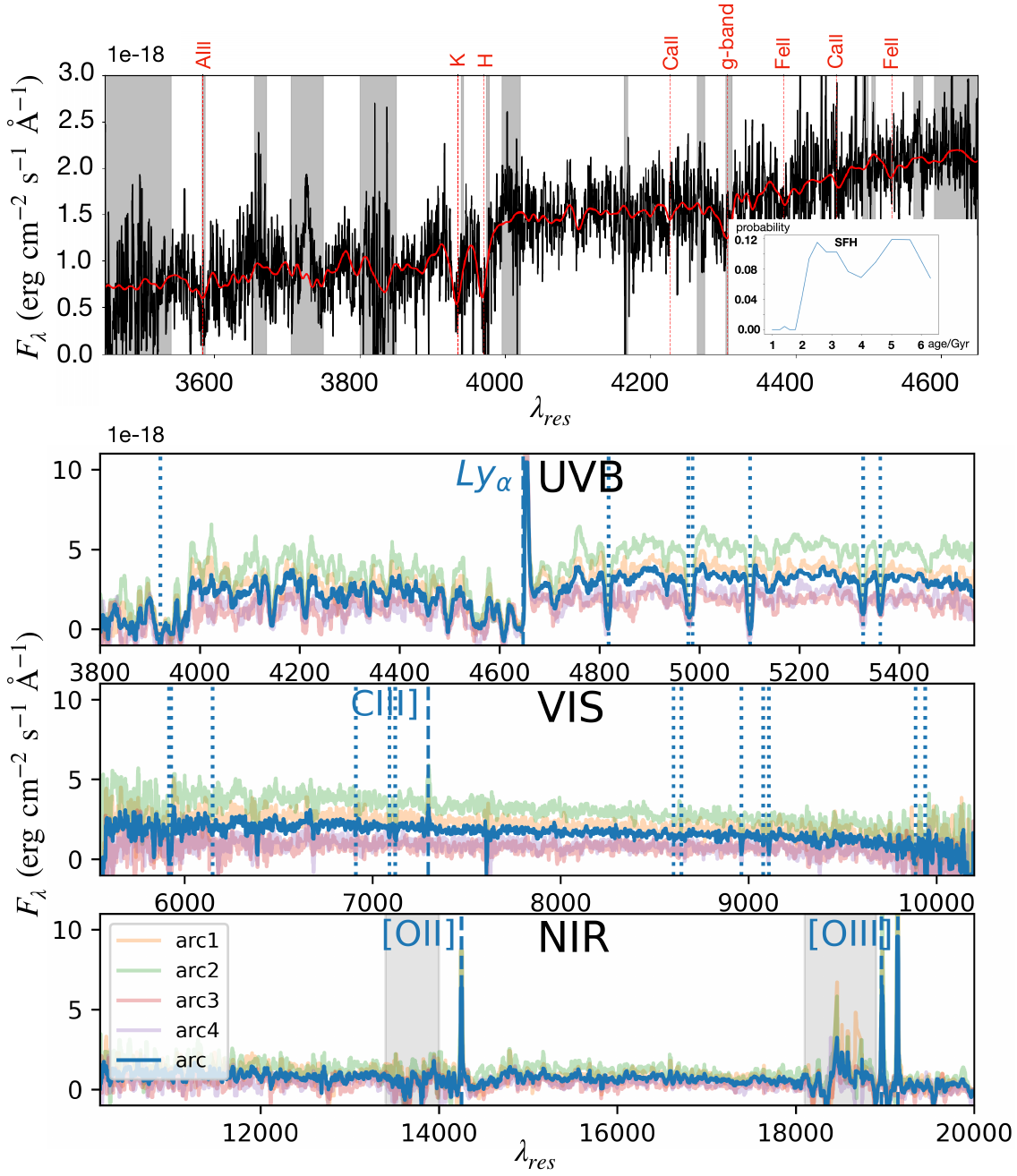}
\caption{Top: optical arm of the  spectra of the deflector (in rest-frame), obtained from X-Shooter. The 4000$\rm \dot{A}$ break can be seen. The velocity dispersion can be inferred from the two characteristic absorption lines, e.g. [K] and [H] bands, of elliptical galaxies. Bottom: the spectra of the lensed source in the NUV, VIS and NIR arms. The four arcs are decomposed from the two slits of different orientations (see Fig. \ref{fig:multi_band_images}). The blue (arc-ave) shows the average flux of the four arcs. The Ly$\alpha$, $\rm O \uppercase\expandafter{\romannumeral2}$, $\rm O \uppercase\expandafter{\romannumeral3}$ lines 
demonstrate that the background source is a Ly$\alpha$ emitter at $z_\text{s}=2.823$.}
\label{fig:spectra}
\end{figure}



KIDSJ0233$-$3447 (RA = 02:33:41.29, Dec = $-$34:47:59.24) was discovered in the southern KiDS footprint through a machine-learning search (\citealt{Li_DR5lens_2021}). In that study, we trained a supervised convolutional neural network based on the ResNet (\citealt{He2016cvpr.confE...1H}) architecture, using simulated strong lenses as positive examples and real KiDS galaxies as negative examples. Applied to the full KiDS DR5 data set ($\sim$350 $\rm deg^2$), the classifier identified 97 high-quality lens candidates, of which KIDSJ0233$-$3447 is particularly noteworthy.
The $g, r, i$ colour image of this system shows two bright blue arcs that form an almost perfect Einstein ring (see Fig. \ref{fig:multi_band_images}). However, no central galaxy is visible in these optical observations. For the same system, the 5-band Near-Infrared ($Z, Y, J, H, Ks$) images from the VIKING Survey (\citealt{Edge2013Msngr_VIKING}) show a small-sized galaxy in the center of the ring.

The optical to near-infrared spectroscopy was collected using the X-Shooter spectrograph at the VLT (ID: 107.22S8, PI: Napolitano). The observations were made in the long-slit mode, with an average seeing of approximately $0.9''$ and airmass of $\sim1.4$. The length of the slits is $11''$, and the widths for the lens are $1.3''$, $1.2''$, $1.2''$ in UVB, VIS, and NIR arms, respectively, while for the arcs are $1.3''$, $1.5''$, $1.5''$. The total integration time was 2h30m for the lens (3 exposures of 3000s each, orange dashed rectangle in Fig. \ref{fig:multi_band_images}) and 1h15m for the arcs (2 exposures of 2250s each, orange solid rectangle). The two different slit orientations served to minimize the contamination of the arcs to the lens and vice versa.

Following the same method as in \citet{Spiniello2019MNRAS.485.5086S}, for the spectra observed in each slit and each arm, we decomposed the wavelength calibrated, background-subtracted and telluric-corrected 2D spectra into three components, two arcs and a central galaxy, using the accurate astrometric information provided by the strong lensing model performed on AO images (see Sec. \ref{sec:lens_model}). The VIS-arm spectrum of the central galaxy is shown in Fig. \ref{fig:spectra}. 
We fitted the spectrum of the central galaxy
with \textsc{pPXF} (\citealt{Cappellari2017MNRAS.466..798C}) and found it to follow an elliptical galaxy template at $z_l = 0.9906\pm0.0001$, showing evident features of absorption lines of [K], [H] and G-band. We have estimated a stellar velocity dispersion inside the slit (VIS, $1.2''$ width, orange slit in Fig. \ref{fig:multi_band_images}) of $309\pm35$ km/s, using the region surrounding the prominent [K] and [H] lines. The fitting also shows a double peaks star formation history (SFH) at the age of $\rm \sim 3 Gyr$ and $\rm \sim 6 Gyr$ (see the bottom right corner in the first row). 
We confirmed the lensing nature of the images by comparing the four spectra
decomposed in the two slits (arc1, arc2, arc3, and arc4, see, in particular, the UVB in Fig. \ref{fig:spectra}), as they show a very similar continuum with the same absorption/emission lines (e.g. Ly$\alpha$, C[\rm \uppercase\expandafter{\romannumeral3}], [O\rm \uppercase\expandafter{\romannumeral2}] and O[\rm \uppercase\expandafter{\romannumeral3}] emissions).
Finally, we determined the source redshift from these lines in the mean spectrum of the four components
and it is $z_s=2.823\pm0.001$.

\subsection{Lensing model with AO image}
\label{sec:lens_model}
The Y-band AO images were obtained with HAWKI@VLT under the same ESO program and were reduced by the HAWK-I Pipeline\footnote{https://www.eso.org/sci/facilities/paranal/instruments/hawki/doc.html}.
The final processed stacked image has a pixel scale of $0.1065''$/pixel and point spread function (PSF) of $0.46''$ (full width at half maximum, FWHM), with a deeper exposure than those from KiDS and VIKING, making it ideal for accurate lensing modelling.


The modeling was performed using the \textsc{Lensed} code (\citealt{Tessore_2016MNRAS.463.3115T_lensed}), which allows for simultaneous modeling of the mass and light distributions. For the light distribution of the lens galaxy, we tested both a de Vaucouleur profile (\citealt{deVaucouleurs1948AnAp...11..247D}) and a more flexible S{\'e}rsic profile (\citealt{Sersic1963BAAA....6...41S}). Both profiles provide an excellent fit to the lens galaxy's light. However, the choice significantly impacts the derived effective radius ($R_e$). We found that the S{\'e}rsic model yields a considerably larger effective radius compared to the de Vaucouleur model (see Table~\ref{tab:model}). This leads to a much higher estimate for the dark matter fraction within $R_e$ in our following analysis, as this fraction is sensitive to the radius of measurement. 
A primary goal of this work is to directly compare the scaling relations of KiDSJ0233-3447 with SLACS (\citealt{Bolton2008ApJ...682..964B}) and SL2S (\citealt{Sonnenfeld2013ApJ...777...98S}) lens samples, as well as with simulations (\citealt{Nipoti2009ApJ...703.1531N}), which predominantly used the de Vaucouleur profile for consistency. Using S{\'e}rsic-derived parameters would introduce a systematic bias. Therefore, to ensure a robust comparison, we adopt the de Vaucouleur profile as our fiducial choice for the main analysis. For full transparency, we report the key results from both light models in our analysis on AO image (see Table~\ref{tab:model}).
Since the source is a Ly$\alpha$ emitter and very little is known about the structure of these galaxies, we decided to use a S{\'e}rsic profile for the light distribution.
Finally, we modeled the total mass distribution of the deflector using a singular isothermal ellipsoid (SIE; \citealt{Kormann1994A&A...284..285K}) profile. The SIE model was selected for its ability to accurately recover the total mass enclosed within the Einstein radius. This total mass measurement is subsequently employed in the combined lensing and dynamical analysis to derive the total mass density slope (see Section \ref{sec:lens_dynamic}).
In the model process, the centers of the mass and light distributions were treated as independent parameters. The best-fit results show they are coincident to within 0.06 arcseconds, which is only slightly larger than half a pixel in the AO imaging. We also tested for the presence of an external shear component but found it to be negligible ( $\gamma_{shear}<0.02$) and its inclusion did not impact our final results. We therefore did not include it in our final model.

\begin{figure}
    \centering
    \includegraphics[width=6.7cm]{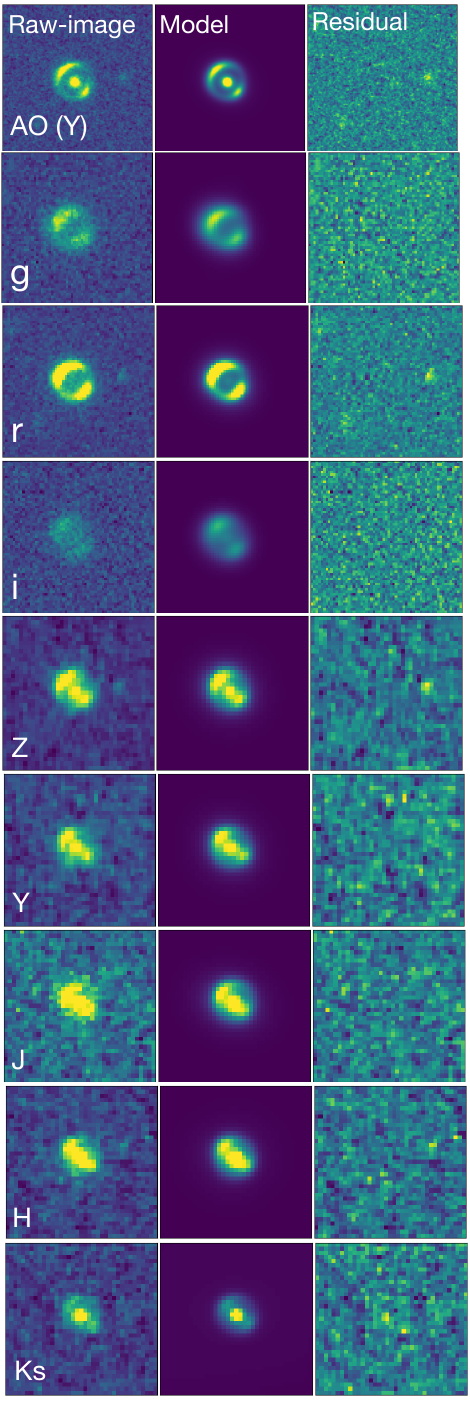}\vspace{3pt}
    \caption{Multi-band strong lensing model of KIDSJ0233-3447. From top to bottom, we show the modeling of the AO (Y-band), three optical ($g,~ r, ~i$) and five NIR (Z, Y, J, H, Ks) images. In each band, the left, middle and right panels show the observed data, the lens model and the residuals, respectively.
    }
    \label{fig:model}
\end{figure}

In the first row of Fig. \ref{fig:model}, we show the Y-band AO image, the best-fitting model, and the residuals obtained by \textsc{Lensed}. The main parameters obtained from the lens modeling are listed in Tab. \ref{tab:model}. The Einstein radius is $R_{\rm Ein}=10.47\pm0.06$ kpc, corresponding to an Einstein mass of $M_{\rm Ein}=7.213\pm0.089 \times 10^{11} M_{\sun}$. The axis ratio for the total lensing mass is $0.932\pm0.008$, which is a little rounder than the axis ratio of the light, $b/a=0.85\pm 0.08$, but 
consistent 
with this latter within the 1 $\sigma$ error. The effective radius of the deflector is $R_{\textrm{eff}}=3.56\pm0.46$ kpc, about 1/3 of $R_{\rm Ein}$. These properties imply that the deflector is a massive elliptical galaxy with a round shape. The effective radius and the S{\'ersic} index $n$ of the source are $R_{\textrm{eff}}=0.991\pm0.184$ kpc and $n=1.34\pm0.27$, respectively, which are typical values for Ly$\alpha$ emitters at $z\sim3$ (e.g., \citealt{Shu2016ApJ...824...86S, Shu2016ApJ...833..264S}).

In order to check the robustness of the lens modeling, we have conducted an additional round of modeling and its subsequent analysis (including Sec. \ref{sec:lens_dynamic} and \ref{sec:stellar_pop}), employing a S{\'ersic} profile for the central foreground galaxy. The parameters obtained from this analysis are also listed in Table \ref{tab:model}. Upon comparison, it is observed that the parameters derived from the assumption of a S{\'ersic} central galaxy are in concordance with those obtained from the de Vaucouleur model, with the exception of the effective radius $\rm R_{eff}$. This discrepancy in $\rm R_{eff}$ subsequently influences the estimation of the total mass and the fraction of dark matter enclosed within the effective radius. Within the scope of this study, we will concentrate on the de Vaucouleur model for our foreground central galaxy assumption. Consequently, we will not delve into an in-depth discussion of the outcomes associated with the S{\'ersic} central galaxy hypothesis. 

\begin{deluxetable}{lrr}
\tablecaption{Summary of the key parameters of the lens system \label{tab:model}}
\tablewidth{0pt}
\tablehead{
Parameters  & de Vaucouleur& S{\'e}rsic
}
\startdata
\hline
\multicolumn{3}{c}{Xshooter spectroscopy}\\ 
\hline
$z_{\rm l}$ & $0.9906\pm0.0001$&$0.9906\pm0.0001$\\
$z_{\rm s}$ & $2.8230\pm0.0001$&$2.8230\pm0.0001$ \\
$\sigma_{\rm slit}$ (km~s$^{-1}$) & $309\pm35$ &$309\pm35$\\
$\sigma_{\rm R_{\rm e}}$ (km~s$^{-1}$) &$343\pm37$& $335\pm37$\\
\hline
\multicolumn{3}{c}{Lensing model(based on AO image)}\\ 
\hline
$R_{\rm Ein}$ (arcsec) & $1.2926\pm0.008$& $1.2899\pm0.006$\\
$R_{\rm Ein}$ (kpc)& $10.47\pm0.06$& $10.45\pm0.05$\\
lens mass $b/a$ & $0.932\pm0.008$& $0.933\pm0.007$ \\
lens mass $PA$ & $141.1\pm3.8$& $141.0\pm3.9$ \\
lens light $n-$index & $4.0$\ (fixed)& $5.60\pm0.78$\\
lens light $b/a$ & $0.85\pm 0.08$ & $0.85\pm 0.08$ \\
lens light $PA$ & $122.6\pm 59.8$& $117.8\pm 58.8$\\
lens light $R_{\rm eff}$ (arcsec)& $0.4392\pm 0.057$ & $0.6206\pm 0.1252$  \\
lens light $R_{\rm eff}$ (kpc) & $3.56\pm 0.46$ & $5.03\pm 1.01$\\
source $n-$index & $1.42\pm0.26$& $1.34\pm0.27$\\
source  $b/a$ & $0.76\pm 0.09$ & $0.83\pm 0.09$\\
source  $R_{\rm eff}$ (arcsec) & $0.123\pm 0.017$& $0.124\pm 0.023$\\
source  $R_{\rm eff}$ (kpc) & $0.986\pm 0.136$& $0.991\pm 0.184$ \\
\hline
\multicolumn{3}{c}{Dynamic and Mass estimates}\\ 
\hline
$\gamma_{tot}$ & $2.195^{+0.086}_{-0.097}$& $2.202^{+0.093}_{-0.106}$\\
$M_{\rm Ein}/(10^{11}M_\sun)$& $7.183\pm0.067$  & $7.213\pm0.089$ \\
$M^{\rm eff}_{tot}/(10^{11}M_\sun)$ & $2.848\pm0.369$& $3.805\pm0.717$ \\
$M_{*}/(10^{11}M_\sun)$ (Chab) & $2.31\pm0.20$& $2.06\pm0.23$ \\
$M_{*}/(10^{11}M_\sun)$ (Salp) & $3.95\pm0.35$& $3.71\pm0.38$ \\
$f_{\rm DM}(R_{\rm eff})$ (Chab) & $0.594\pm0.088$& $0.729\pm0.081$ \\
$f_{\rm DM}(R_{\rm eff})$ (Salp) & $0.307\pm0.151$ & $0.512\pm0.141$\\
\hline
\enddata
\tablecomments{The key parameters of the lens system. The first line specifies the lens light in the modelling. X-shooter spectroscopy: redshift of the lens ($z_l$) and source ($z_s$) and the velocity dispersion of the lens calculated in the slit and at $R_{\rm eff}$. Lensing models from AO image: Einstein radius ($R_{\rm Ein}$) in arcsec and kpc, followed by self-explaining parameters related to the total mass (labeled by `lens mass'), lens light profile (labeled by `lens light') and source light profile (labeled by `source'). Dynamic and Mass estimates: summary of the total mass density slope ($\gamma$), mass estimates from the lensing model ($M_{\rm Ein}$), and stellar population $M_*^{SED}$ (see text for details) together with the projected DM fractions ($f_{\rm DM}$).}
\end{deluxetable}

\subsection{Joint lensing and dynamical analysis}
\label{sec:lens_dynamic}
By combining the ray tracing lens modelling and the dynamics of the central galaxy, we have enough constraints to infer more detailed properties of the total mass of the lensing system, e.g. its central slope (see e.g. \citealt{Treu2002ApJ...575...87T}).
For this purpose, we have assumed that the density distribution of the lens galaxy in three dimensions follows a power-law profile of the form
\begin{equation}
    \rho=\rho_{0}(\frac{r}{r_0})^{-\gamma}
    \label{eq:rho}
\end{equation}
where $\rho_0$ and $r_0$, jointly, can be determined by the Einstein mass and Einstein radius, and $\gamma$ denotes the logarithmic slope. We then solve the spherical Jeans equation (\citealt{Binney1987gady.book.....B})
\begin{equation}
\frac{1}{\nu} \frac{d(\nu \bar{v_r ^2})}{dr}+2 \frac{\beta \bar{v_r ^2}}{r}=-\frac{GM(<r)}{r^2},
\label{Equ:Jeans_equation}
\end{equation}
where $\nu$ represents the density of the light particles
in the galaxy, assumed to be proportional to the three-dimensional light distribution. The anisotropy parameter, $\beta$ (\citealt{Osipkov1979PAZh....5...77O}; \citealt{Merritt1985AJ.....90.1027M}), is set to be 0 (isotropic orbits) in our analysis, consistent with many works that analyze the velocity dispersion of massive ETGs (e.g., \citealt{Koopmans2009ApJ...703L..51K, Auger2009ApJ...705.1099A}). Finally, $M(<r)$ denotes the total mass contained within a sphere of radius $r$, corresponding to the density in Eq. \ref{eq:rho}.
We performed the lensing and dynamical analyses following the method described in \cite{Li2018MNRAS.480..431L} (see also \citealt{Koopmans2006ApJ...649..599K, Koopmans2009ApJ...703L..51K, Bolton2012ApJ...757...82B, Sonnenfeld2013ApJ...777...98S}). Briefly, the method consists of minimizing the $\chi^2$ between the velocity dispersion obtained by projecting the solution of Eq. \ref{Equ:Jeans_equation} within an aperture corresponding to the effective radius, and the observed one within the same aperture. At each iteration of the $\chi^2$, the $M_{\rm Ein}$ and the $R_{\rm Ein}$ are used to derive the $\rho_0$ in Eq. \ref{eq:rho} used to solve Eq. \ref{Equ:Jeans_equation}. The atmospheric seeing and the aperture luminosity weighting function are also considered (see detail in \citealt{Schwab2010ApJ...708..750S}). 
We found the best-fit total mass density slope
to be $\gamma_{\rm tot}=2.14^{+0.06}_{-0.07}$. Assuming a power-law mass distribution, the total mass within the $R_{\rm eff}$ (consisting of both stellar and dark matter) was determined to be $M^{\rm eff}_{\rm tot}=(2.85\pm0.37)\times 10^{11} M_\sun$.

\subsection{Stellar population from 8-band ray tracing model}
\label{sec:stellar_pop}
We took advantage of the optical plus NIR photometry from KiDS and VIKING to derive the stellar population parameters of the lens, given that the signal-to-noise ratio of the X-shooter spectra is insufficient to perform a detailed study of the spectral indices. We have performed the lensing model on 8-bands ($g r i Z Y J H K_{\rm s}$) images to derive accurate and uncontaminated multi-band magnitudes for the central galaxy. The light of the central galaxy is fitted with a de Vaucouleur profile, while the mass is assumed to be SIE. Details about the modeling can be found in appendix \ref{sec:appendix}. We show the model and residuals in Fig. \ref{fig:model} for a sample of KiDS and VIKING bands, including the $Y$-band to confront the AO model.

We performed the SED fitting with the multi-band photometry
using the \textsc{Cigale} code. For this analysis we have assumed \citet[][BC03]{Bruzual2003MNRAS.344.1000B} stellar population templates. Parameters, such as the metallicity, extinction, star-forming rate, and age of the main stellar population were set free to vary. We used both Chabrier and Salpeter IMF, which returned a total stellar mass of  $M_{\rm SED}^{\rm Chab}=(2.31\pm 0.20)\times10^{11}M_\odot$, and $M_{\rm SED}^{\rm Salp}=(3.95\pm 0.35)\times10^{11}M_\odot$, respectively. Given $R_{\rm eff}=3.56\pm0.46$ kpc, we found that the mass-size relation of this galaxy is fully consistent with ETGs at redshift $z\sim1$ (see \citealt{Sonnenfeld2013ApJ...777...98S}). Finally, by subtracting the stellar mass from the total mass, we obtained the DM mass, and then the DM fraction inside the effective radius, which are $f_{DM}^{Chab}=0.594\pm0.088$ for the Chabrier and $f_{DM}^{\rm Salp}=0.307\pm0.151$ for the Salpeter IMF.

The stellar population analysis returned a galaxy age of $5.9$ (Chabrier IMF) or $5.6$ Gyrs (Salpeter IMF), suggesting an old stellar population and an early formation epoch
but with a rather extended delayed star formation history SFH$\propto t/\tau^2\exp(-t/\tau)$, with a $\tau\sim 0.8$ Gyr.
The old age has been confirmed by the \textsc{pPXF} analysis of the X-Shooter spectra showing a double-peaked SFH, at $\sim5.5$ Gyr and $\sim2.5$ Gyr in lookback time (see the top panel in Fig. \ref{fig:spectra}), hence fully compatible with old age and an extended SFH. This is a much older age than the one found from another high-z lens, MG 2016+112
which shows younger stellar populations (\citealt{Treu2002ApJ...575...87T}).

\section{Discussion on the properties}
\subsection{Total mass density slope and dark matter fraction}
\label{sec:dis_density_profile}

As mentioned above, the lensing galaxy KiDSJ0233-3447 is the second galaxy at $z_l\sim 1$ for which a total density slope measurement has been obtained after MG\,2016+112. The latter system has many similarities with KiDSJ0233-3447, such as the lens and source redshift, and the velocity dispersion inside the effective radius. The exact values are listed below:
\begin{itemize}
    \item \textbf{KiDSJ0233-3447:} $z_l = 0.9906$, $z_s = 2.823$, $\sigma_{R_e} = 343 \pm 37$\,km/s, and $\gamma_{\text{tot}} = 2.14^{+0.06}_{-0.07}$.
    \item \textbf{MG\,2016+112:} $z_l = 1.004$, $z_s = 3.263$, $\sigma_{R_e} = 304 \pm 27$\,km/s, and $\gamma_{\text{tot}} = 2.0 \pm 0.1$.
\end{itemize}
While the total density slopes, $\gamma_{\text{tot}}$, are consistent with each other within their $1\sigma$ uncertainties, the value for KiDSJ0233-3447 is notably steeper and deviates from a purely isothermal profile ($\gamma_{\text{tot}} = 2$) at the $2\sigma$ level. In contrast, the slope of MG\,2016+112 is perfectly consistent with an isothermal distribution.
This difference in $\gamma_{\text{tot}}$ likely reflects the distinct evolutionary histories of the two galaxies. The steeper slope in KiDSJ0233-3447 suggests a more significant baryonic concentration at its center, which in turn induces a stronger adiabatic contraction of the dark matter halo compared to MG\,2016+112. We present a detailed analysis of this scenario, which connects the stellar IMF to the halo contraction, in Section~\ref{sec:imf_dm_slope}.

\cite{Ruff2011ApJ...727...96R} and \cite{Bolton2012ApJ...757...82B} claimed that the total mass density slope $\gamma_{tot}$ should increase with cosmic time. However, this appears inconsistent with simulations of dry merger-driven evolution in ETGs. \cite{Nipoti2009ApJ...703.1531N} suggested that dry mergers might preserve the nearly isothermal structure of galaxies during $0<z<1$. Later analyses by \cite{Sonnenfeld2014ApJ...786...89S} demonstrate that pure dry mergers predict a substantial decrease in density slope over time, conflicting with lensing observations indicating either increasing or constant $\gamma_{tot}$. When incorporating modest dissipative processes (e.g., gas accretion), the central density profile can steepen, offsetting the flattening effects of dry mergers. This mechanism enables $\gamma_{tot}$ to remain constant from z$\sim$1 to z=0. 

Given the scarcity of gas in ETGs to allow wet mergers, the only viable reconciliation between observations and simulations requires non-evolution of $\gamma_{tot}$ with redshift. SL2S lenses ($0\lessapprox z\lessapprox0.8$) measured $\gamma_{tot}=2.05\pm0.06$ (\citealt{Sonnenfeld2013ApJ...777...98S}). While SLACS samples previously suggested slightly steeper slopes ($\gamma_{tot}=2.078\pm0.027$), recent work by \cite{Sonnenfeld2024A&A...690A.325S} reveals these measurements may be overestimated by $\sim 0.1$ due to selection biases. Lenses at $z\sim1$ like KiDS0233-3447 and MG 2016+112 exhibit slopes comparable to these lower-redshift lenses. We therefore suggest that the "blue eye" shows no systematic deviation from low-redshift slopes, supporting slope stability at $z<1$ (e.g., \citealt{Koopmans2006ApJ...649..599K}; \citealt{Sonnenfeld2013ApJ...777...98S}). This evidence suggests ETG evolution through combined dry mergers and moderate dissipationless accretion. Such hybrid evolution simultaneously explains observed size growth and $\gamma_{tot}$ stability.

\begin{figure*}
\centering
\includegraphics[width=18cm]{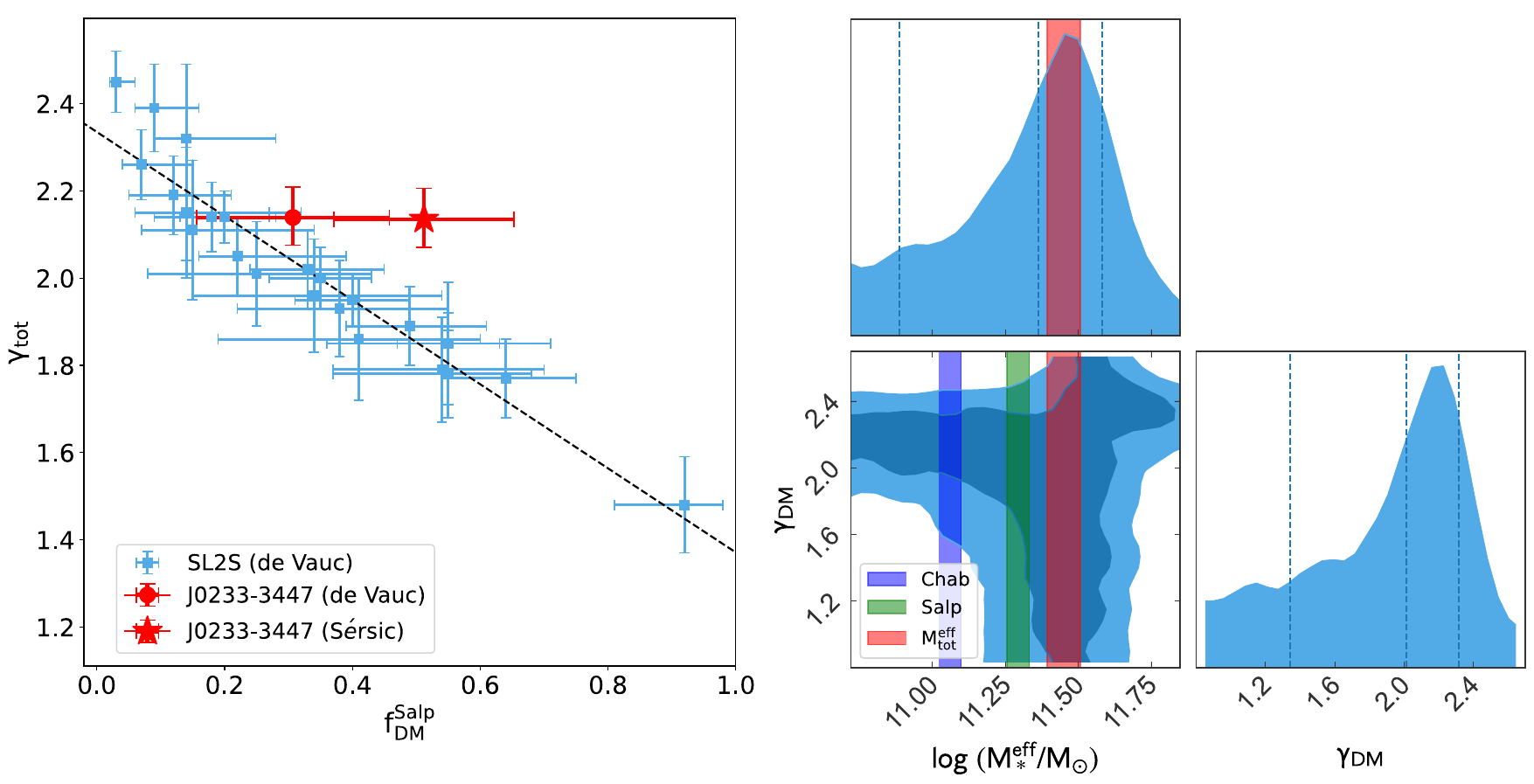}
\caption{Left: The relationship between total mass density slope $\gamma_{tot}$ and dark matter fraction (Salpeter IMF, $f_{DM}^{\rm Salp}$) for SL2S sample (squares) and of KiDSJ0233-3447 by assuming de Vaucouleur (dot) and S{\'e}rsic (star) foreground light profile. The black dashed line shows the line fit of the SL2S sample. Right: The MCMC inference for dark matter slope $\gamma_{DM}$ and the stellar mass inside the $\rm R_{eff}$. The stellar mass inferred from Chabrier and Salpeter IMF (also inside the $\rm R_{eff}$) are shown with blue and green bars, respectively.} 
\label{fig:DM_gamma}
\end{figure*}

However, the total density slope $\gamma_{tot}$ is a combination of 1) the shallower (typically $\sim 1$) dark matter and the steeper (typically $\sim 3$) stellar matter slopes (see e.g. \citealt{2006AJ....132.2701G}) in the galaxy centers and 2) their relative mass fraction. A larger DM fraction generally leads to a shallower total density profile, closer to the DM slope, while a lower DM fraction would produce a slope more biased toward the steeper starlight. 
Interestingly, when DM and stars almost balance, the slope conspires toward an isothermal value, $\gamma_{tot}\sim2$. Lens systems, indeed show a variety of slopes and a correlation with $f_{\rm DM}$ that follows this intuitive picture, as it can be seen
in the left panel of Fig. \ref{fig:DM_gamma}. Here, we show the correlation between the density slope $\gamma_{tot}$ and the dark matter fraction $f_{\rm DM}^{\rm Salp}$ for the SL2S sample from \cite{Sonnenfeld2013ApJ...777...98S}. 
We have chosen the SL2S sample for uniformity on the model assumptions, as they provide the $f_{\rm DM}$ inside the effective radius $R_{\rm eff}$, like us, and unlike SLACS (\citealt{Auger2009ApJ...705.1099A}), using half $R_{\rm eff}$.   
Also, SL2S employs multi-band photometry to calculate the stellar mass, although they have a narrower wavelength baseline than ours. Finally, the SL2S sample covers a wider redshift range than the SLACS, with the highest-$z$ lens at $z\sim 0.781$, closer to the redshift of the Einstein ``blue eye''. Finally we use the de Vaucouleur profile as in SL2S.

In Fig. \ref{fig:DM_gamma}, KiDSJ0233-3447 shows a $\gamma_{tot}=2.14^{+0.06}_{-0.07}$, which is fully consistent with the correlation from the SL2S sample. In the same figure, we also show that the situation is different if we assume a S\'ersic model for the foreground galaxy instead of a de Vaucouleur (full lensing parameters reported in Table \ref{tab:model}), mainly due to the larger $R_{\rm eff}$.
We might expect that a similar correction toward larger DM fraction could apply to the SL2S sample if a S\'ersic model were adopted but, overall, the conclusion about the consistent trend between $f_{\rm DM}$ and $\gamma$ would remain unchanged, not certainly worsened.

Here we note that the observed $\gamma_{tot}-f_{DM}$ relation in the SL2S sample, as well as in other lensing samples (e.g., SLACS; \citealt{Shajib2021MNRAS.503.2380S}), exhibits discrepancies with cosmological simulations. Hydrodynamical simulations either reproduce the 
$\gamma_{tot}$ distribution but systematically underestimate $f_{DM}$ when baryonic physics is minimally included (e.g., \citealt{Duffy2010MNRAS.405.2161D, Johansson2012ApJ...754..115J}), or recover the observed $f_{DM}$ with strong feedback at the expense of predicting shallower slopes than observed (e.g., \citealt{Dubois2013MNRAS.433.3297D, Xu2017MNRAS.469.1824X}). The $\gamma_{tot}-f_{DM}$
relation of KIDSJ0233-3447 aligns with previous lensing studies at lower redshifts (e.g., \citealt{Sonnenfeld2013ApJ...777...97S, Shajib2021MNRAS.503.2380S}), indicating the tension with simulations is also exists at higher redshift ($z\sim 1$). This persistent mismatch highlights the need for either refined implementations of feedback mechanisms in cosmological simulations or a critical re-examination of potential systematic biases in observational analyses.

\subsection{IMF and dark matter slope}
\label{sec:imf_dm_slope}
The analysis presented in Section \ref{sec:dis_density_profile} adopts the Salpeter IMF as a wavelength baseline for comparison, aligning with the approach utilized in the SL2S study. It is acknowledged, however, that the IMF is not necessarily confined to the Salpeter IMF, with evidence suggesting variations in the IMF across different galaxies. The choice of IMF influences the estimation of the dark matter fraction, which, in turn, has implications for the measurement of the dark matter density slope. Specifically, the possibility that KiDSJ0233-3447 favors a lighter or a heavier IMF, for a fixed total slope, has direct consequences on the inferred DM halo density slope. For instance, a higher DM fraction given from a lighter IMF would imply an excess of DM which might not be compatible with some standard $\Lambda$CDM halo. 
Hence, we need to evaluate how much the DM slope can vary in response to some physically motivated IMF variation. 

To check this, we have combined lensing and dynamics again. But instead of a single power-law, we consider the total mass of the galaxy in Eq. \ref{Equ:Jeans_equation} made of the stellar and the dark matter components, and then use Bayesian inference to obtain the posterior distribution of dynamical stellar mass and dark matter slope.
The stellar mass is assumed to follow the de Vaucouleur profile, as in Table \ref{tab:model}, with a dynamical mass-to-light ratio as a free parameter ($M_*^{\rm dyn}(r)=M/L_*^{\rm dyn}\times L_{\rm tot}^{\rm deV}(r)$, being the $ L_{\rm tot}^{\rm deV}(r)$ the growth curve of the de-projected de Vaucouleur profile). Here, we continue to employ the de Vaucouleur profile to ensure a robust and direct comparison with key findings in the literature (e.g., \citealt{Nipoti2009ApJ...703.1531N}; \citealt{Sonnenfeld2013ApJ...777...98S}), as has been demonstrated in Section~\ref{sec:lens_model}. Adopting a S{\'e}rsic profile would systematically alter the derived dark matter fraction, biasing any such comparison and thus affecting our conclusions.
For the dark matter component, we adopt a power-law density profile, $\rho_{\rm DM}(r) \propto r^{-\gamma_{\rm DM}}$ (e.g., \citealt{Dutton2014MNRAS.438.3594D, Sonnenfeld2015ApJ...800...94S}). This allows for a direct comparison with the standard Navarro-Frenk-White (NFW) profile, which features an inner slope of $\gamma_{\rm DM}=1$ and provides a good description of dark matter halos in cosmological N-body simulations (\citealt{Navarro1997ApJ...490..493N}). To probe how the slope might change with a physically motivated IMF variation, we treat  $\gamma_{\rm DM}$ as a free parameter in our Bayesian inference.
Finally, we note that the stellar mass obtained here is dynamically inferred and we can use its ratio with the SED/IMF inferred stellar masses, as a proxy of the variation of the IMF.

In Fig. \ref{fig:DM_gamma}, we report the confidence contours (68\% and 95\%) and the marginalized posterior probabilities (16\%, 50\% and 84\%, from left to right vertical dashed lines) of the two parameters in our model. To have a more physical insight of the inferences, on the x-axis we report the stellar mass inside the $R_{\rm eff}$ corresponding to the inferred $M/L_*^{\rm dyn}$ as this can be compared with the inferences of the SED/IMFs (namely Chabrier and Salpeter IMF) and with the total mass inside the same radius from the lensing and dynamics, all marked as vertical stripes in the figure. This latter value, provides a physical upper limit for the stellar mass, as the mass in stars cannot exceed the total enclosed mass. It is plotted for a posteriori comparison and helps to interpret the physically region of the parameter space.

From the confidence contours, we see a degeneracy between $M_*^{\rm eff}$ and $\gamma_{DM}$. IMF normalizations above a Salpeter IMF are compatible with a larger variety of DM slope, between $\gamma_{\rm DM}\sim2.3$ and down to $\gamma_{\rm DM}\sim1$. However, using the $M_{\rm tot}^{\rm eff}$ as a physical upper limit for the stellar mass, we see that the accessible parameter space at low $\gamma_{\rm DM}$ is strongly restricted. 
A reasonable solution for an IMF steeper than the Salpeter IMF
should have a stellar mass $\log M_*^{\rm eff}=11.36\pm0.05$ (i.e. around the lower limit of $M_{\rm tot}^{\rm eff}$) and a $\gamma_{\rm DM}\sim1.6\pm0.6$, which would correspond to a $f_{\rm DM}=0.20$, perfectly reconciling the ``blue eye'' with the SL2S sample (see \S\ref{sec:dis_density_profile}). 
This case is statistically consistent with a standard NFW profile, because the NFW inner slope of 
$\gamma_{DM}$=1 lies within the $1\sigma$ confidence interval of our measurement, which is $\gamma_{\rm DM}\sim1.6\pm0.6$.

On the other hand, if one wants to stick to a Salpeter IMF, the $\log M_*^{\rm eff}=11.30\pm0.04$ and a $\gamma_{\rm DM}\sim2.0\pm0.4$, which would correspond to a $f_{\rm DM}=0.31$ (see also the lensing and dynamics analysis presented in \S\ref{sec:dis_density_profile}).  Considering that the DM slope is estimated within the $R_{\rm eff}$, we can follow \citet{Napolitano2010MNRAS.405.2351N} (see also \citealt{2006AJ....132.2701G}) and assume a dilution of the dark matter slope of $\Delta \gamma_{\rm DM}\sim0.3$ with respect to the true central cusp, to find that this case is clearly steeper than a NFW ($\gamma_{\rm DM}^{\rm true}=2.0-0.4>1.0\pm0.3$) and more compatible with a contracted halo. A similar conclusion was also reached by \cite{Sonnenfeld2012ApJ...752..163S}, where they analyzed a double Einstein ring lens at a low redshift ($z=0.222$) and inferred a dark matter slope of $\gamma_{DM}=1.7\pm0.2$ and \citet{Napolitano2010MNRAS.405.2351N} who found $\gamma_{DM}\sim1.6$ for a sample of local galaxies. 
Such a co-existence of Salpeter IMF and dark matter halo contraction has also been found in  other works (e.g., \citealt {Treu2010ApJ...709.1195T, Tortora2010ApJ...721L...1T, Spiniello2011MNRAS.417.3000S, Sonnenfeld2013ApJ...777...98S}),
suggesting it can be a regular feature in massive elliptical systems (but see \citealt{2020ApJ...899...87M}). 

Finally, according to the posterior probability, we cannot rule out the possibility of a lighter IMF than the Salpeter IMF, which would not alter the overall DM slope, yet compatible with a contracted halo. In particular, we see that a Chabrier or even lighter IMF is compatible with a steep DM slope up to $\gamma_{\rm DM}\gsim2.0$, which is strongly deviating from the $\Lambda$CDM DM-only simulation predictions (i.e. a NFW profile, \citealt{Navarro1997ApJ...490..493N}). The main implication of such an IMF would be an increased dark matter fraction of $f_{\rm DM}\sim 0.6$ (see Tab. \ref{tab:model}). Such a DM fraction is larger than typical values found in observations for lower redshift galaxies (e.g. COMA, \citealt{Thomas2007MNRAS.382..657T}; SLACS ($f_{DM}<0.6$ \citealt{Barnabe2011MNRAS.415.2215B}), and almost twice that expected from simulations for a Chabrier IMF at z$\sim$1 (\citealt{ Oser2010ApJ...725.2312O, Remus2017MNRAS.464.3742R}).

Although the arguments discussed above support both the Salpeter and Chabrier IMFs with a contracted halo, we can provide more direct evidence to support that KiDSJ0233-3447 favors a heavier IMF (e.g., Salpeter) combined with contraction over a standard uncontracted NFW profile.
Indeed, in a heavy halo contraction, one expects the dissipation of cold gas to play a strong role (see e.g. \citealt{2004ApJ...616...16G, 2005ApJ...634...70S}). According to the inferred SFH, both from the multiband photometry and the X-Shooter spectrum, the ``blue eye'' central galaxy has gone through a quite long and massive star-formation activity, which lasted for $\sim 6$ Gyr and ended just before $z\sim1$. In particular, the double-peaked SFH suggested by \textsc{pPXF} might even suggest a major accretion event. 
We argue that this extended SFH might have favored the halo contraction of the central galaxy. We remark here that an extended (even double-peaked) SFH is consistent with the two-stage formation model proposed by \citet{2013MNRAS.435.2274W} that can explain both a bottom-heavy IMF and a solar metallicity observed in high-$z$ ETGs. According to this model, galaxies have formed $\sim$10\% of their stars during a very short burst, lasting
0.3 Gyr, with a top-heavy IMF, then followed by a second stage where the bulk of the stellar mass is formed in about 1 Gyr, with a bottom-heavy IMF. If so, the ``blue eye'' central galaxy can be a test bench of this model, showing eventually even a more prolonged process, possibly driven by the higher velocity dispersion (and hence mass) of KIDSJ0233-3447, with respect to the typical galaxies considered in \citet{2013MNRAS.435.2274W}.

\section{conclusion}
We have provided the first full optical + NIR photometric and spectroscopic analysis of an Einstein ring at $z\sim1$. KiDSJ0233-3447 is a strong lens candidate discovered in KiDS in \citealt{Li_DR5lens_2021}, which we followed up with the X-shooter and HAWKI instruments at the VLT. The foreground galaxy is an early-type galaxy (ETG) located at $z_l=0.9906$, while the background source is a Ly$\alpha$ emitter at $z_s=2.832$.


We collected a combined multi-band (optical to infrared), AO, and spectroscopic dataset for this lens, which allowed us to perform a complete multi-band analysis of the system, including an 8-band lens model. With the analysis, we can constrain The lens parameters in the 8-band models with excellent accuracy, despite the different seeing conditions and image quality of the cameras adopted. The Einstein radius and effective radius are $10.47\pm0.06$ kpc and $3.56\pm0.46$ kpc, respectively and the total mass density slope (obtained via lensing-dynamical analysis) is $2.139^{+0.064}_{-0.070}$, which is close to the isothermal density profile, consistent with previous lower-redshift strong lenses. This supports a scenario of dry merger plus modest dissipative processes (such as gas accretion) driving the ETGs evolution between $0<z<1$ (\citealt{Sonnenfeld2014ApJ...786...89S}).


We reported a dark matter fraction within $R_{\rm eff}$ to be $f_{DM}=0.307\pm0.151$ (assuming a Salpeter IMF), which is consistent with values typically observed in strong lensing systems at lower redshifts and similar masses. To explore whether this dark matter fraction could result from an unaccounted heavier IMF, we found that such a scenario would imply a standard NFW halo (i.e., $\gamma_{DM=1}$)  with $f_{\rm DM}\sim 0.20$. On the other hand, if we assume a Salpeter IMF, the observed dark matter fraction would suggest a contracted halo with a dark matter slope of $\gamma_{\rm DM}>1.6$, consistent with other evidence from low-redshift galaxies.

The old age ($\sim6$ Gyr) and the extended star formation history derived by both the 8-band photometry from the ray-tracing and the X-Shooter of the central spectrum, support the contraction + Salpeter IMF. In particular, the Einstein ``blue eye'' provides evidence of the two-stage IMF scenario \citep{2013MNRAS.435.2274W} for early-stage of ETG formation.

Finally, KiDSJ0233-3447 represents a reference case for analyses that will combine similar wide wavelength baseline datasets, from optical to NIR, from future surveys from ground and from space.

\section*{Acknowledgements}
Rui Li acknowledges the National Nature Science Foundation of China (No. 12203050) and the Natural Science Foundation of Henan Province of China (Grant No.242300420235). NRN acknowledges financial support from the Research Fund for International Scholars of the National Science Foundation of China, grant n. 12150710511. Ran Li acknowledges the support of National Nature Science Foundation of China (Nos 11988101), the support by National Key R$\&$D Program of China No. 2022YFF0503403, the support from the Ministry of Science and Technology of China (Nos. 2020SKA0110100),  the science research grants from the China Manned Space Project, CAS Project for Young Scientists in Basic Research (No. YSBR-062), and the support from K.C.Wong Education Foundation. CT acknowledges funding from INAF research grant 2022 LEMON. HCF acknowledge the financial support of the National Natural Science Foundation of China (grant No. 12203096) and Yunnan Fundamental Research Projects (grant No. 202301AT070339). GD acknowledges support by UKRI-STFC grants: ST/T003081/1 and ST/X001857/1

\bibliography{highz_lens}{}
\bibliographystyle{aasjournal}



\begin{appendix}
\setcounter {table} {0} \renewcommand {\thetable} {\Alph {section}\arabic {table}}
\section{Multi-band lens model}
\label{sec:appendix}

In this appendix, we present the details of the multi-band lens modeling that we performed to obtain the source-uncontaminated photometry of the lens in 8 bands: $g r i Z Y J H K_{\rm s}$. We used the same code \textsc{Lensed} and method as described in Section \ref{sec:lens_model}, but applied them to the KiDS and VIKING images instead of the AO image. We modeled the lens light, source light and mass distribution with de Vaucouleur profile, elliptical S{\'e}rsic profile, and SIE profile, respectively. 
For the $g$, $r$, and $i$-band fits, the model had a total of 18 free parameters: 6 for the lens light (x/y center, total magnitude, effective radius, axis ratio, and position angle), 5 for the SIE mass profile (x/y center, Einstein radius, axis ratio, and position angle), and 7 for the source light (x/y center, total magnitude, effective radius, S{\'e}rsic index, axis ratio and position angle). We varied the parameters of these profiles independently in each band.
We used the same code as in Section \ref{sec:lens_model} to explore the parameter space and obtain the best-fitting models and the uncertainties.
For the lower-resolution NIR images, where the light profile parameters are less constrained, we fixed the effective radius and axis ratio of the lens light, as well as the S{\'e}rsic index and axis ratio of the source light, to the best-fit values from the AO model. This reduced the number of free parameters in these fits to 14 (4 for lens light, 5 for lens mass, and 5 for source light).

Table \ref{tb:multi_band_model_para} summarizes the main parameters obtained from the multi-band lens modeling. The total magnitudes of the lens, as well as the effective radii, the axis ratios, and the position angles, are reported in the table. The Einstein radii, axis ratios, and position angles of the mass profiles are also shown. Fig. \ref{fig:model} shows the images, the models, and the residuals in each band. The models can reproduce the observed features of the lens system, such as the lensed arcs and the counter-image, with small residuals. The parameters obtained from the multi-band lens modeling are consistent with those obtained from the AO model within the uncertainties. 
Despite leaving the total mass parameters of the lens model free to vary, we have found an incredibly tight distribution of their values over the 8-band models, with the $R_{\rm Ein}=1.28\pm0.02$ arcsec, the $(b/a)_{mass}=0.94\pm0.02$ and the $PA_{mass}=148\pm9$ deg, which are fully consistent with the AO parameters in Tab. \ref{tab:model}. This indicates that the mass distribution of the lens is robust and stable. The effective radii of the light of the central galaxy show some variation between the $g r$ bands and the other bands, which is mainly due to the low signal-to-noise ratio in the $g r$ bands.

\begin{table*}[htbp]
\begin{center}
\caption{Parameters of the multi-band lens model.}
\label{tb:multi_band_model_para}
\begin{tabular}{l l l l l l l l l}
\hline \hline
band  & g & r & i & Z & Y & J & H& Ks \\
\hline
\hline
$\rm mag_*$   & 23.82$\pm$0.14 &22.75$\pm$0.21 & 22.13$\pm$0.33 & 21.24$\pm$ 0.01 & 20.68$\pm$0.01 & 20.24$\pm$0.01 & 19.74$\pm$ 0.01& 19.13$\pm$0.01  \\
$\rm R_{eff\ *}$ (arcsec)  & 1.76$\pm$0.19     &1.57$\pm$0.31    & 0.48$\pm$0.32    & 0.26(fixed)    & 0.26(fixed)  & 0.26(fixed)   & 0.26(fixed)     & 0.26(fixed) \\
$\rm (b/a)_*$      &0.52$\pm$0.21      &0.80$\pm$0.13         &0.64$\pm$0.20     & 0.85(fixed)     & 0.85(fixed)    &0.85(fixed)    & 0.85(fixed)     & 0.85(fixed) \\
$\rm PA_*$ (deg)    &78.6$\pm$43.9     &84.0$\pm$42.6    &80.6$\pm$54.0     & 85.6$\pm$0.7    & 48.7$\pm$1.4   &144.8$\pm$0.1  & 76.5$\pm$0.5    & 170.9$\pm$ 0.1\\
$\rm R_{Ein}$ (arcsec)  &1.31$\pm$0.07      &1.30$\pm$0.02 &1.28$\pm$0.19     & 1.28$\pm$0.01     &1.29$\pm$0.01     &1.30$\pm$0.01   &1.27$\pm$0.01      &1.23$\pm$0.01\\
$\rm (b/a)_{mass}$ &0.94$\pm$0.01      &0.95$\pm$0.01         &0.94$\pm$0.02     & 0.95$\pm$0.01     & 0.92$\pm$0.01    &0.90$\pm$0.01   & 0.95$\pm$0.01     & 0.97$\pm$0.01\\
$\rm PA_{mass}$ (deg)&148.5$\pm$4.5    &145.5$\pm$1.3     & 154.5$\pm$8.4    &140.8$\pm$0.1   & 159.13$\pm$0.3   &132.26$\pm$0.1   &142.28$\pm$0.1     & 157.89$\pm$0.1\\
$\rm mag_{sour}$& 23.50$\pm$0.13   &  23.81$\pm$0.08   & 22.79$\pm$0.22   &-5.11$\pm$ 0.01    & -4.48$\pm$0.02 & -5.49$\pm$0.01   &  -5.27$\pm$0.01   & -6.14$\pm$0.01 \\
$\rm R_{eff\ sour}$ (arcsec)& 0.18$\pm$0.02   &  0.09$\pm$0.07  & 0.17$\pm$0.03   & 0.07$\pm$0.01    &  0.11$\pm$0.01  & 0.17$\pm$0.01   & 0.15$\pm$0.01   &0.12$\pm$0.01 \\
$\rm n_{sour}$& 0.93$\pm$0.22         &  1.18$\pm$0.10  & 1.02$\pm$0.32   & 1.42(fixed)   &  1.42(fixed)  & 1.42(fixed)   & 1.42(fixed)   &1.42(fixed) \\
$\rm (b/a)_{sour}$& 0.61$\pm$0.07         &  0.49$\pm$0.03  & 0.84$\pm$0.10   & 0.76(fixed)   &   0.76(fixed)  &  0.76(fixed)  &  0.76(fixed)   &0.76(fixed) \\
$\rm pa_{sour}$ (deg)& 29.7$\pm$5.7         &  14.7$\pm$2.1  & 87.5$\pm$54.8   & 30.4$\pm$0.5   &  163.4$\pm$1.4  & 10.2$\pm$0.1   &  116.3$\pm$0.1  &17.5$\pm$0.1 \\
\hline \hline
\end{tabular}
\end{center}
\textsc{Note.} ---Parameters of the multi-band lens modelling.
The first four rows list the best-fit parameters for the lens light profile: magnitude ($\rm mag_*$), effective radius ($\rm R_{eff\ *}$), axis ratio $\rm (b/a)_*$, and position angle ($\rm PA_*$). Rows 5–7 present the mass model parameters: Einstein radius ($\rm R_{Ein}$), axis ratio $\rm (b/a)_{mass}$, and position angle ($\rm PA_{mass}$). The final rows list the parameters for the lensed source: magnitude ($\rm mag_{sour}$), effective radius ($\rm R_{eff\ sour}$), S{'e}rsic index ($\rm n_{sour}$), axis ratio $\rm (b/a)_{sour}$, and position angle ($\rm pa_{sour}$).
\end{table*}

\end{appendix}

\end{document}